\documentclass[10pt]{iopart}
\usepackage{graphicx}
\begin{document}

\title[Conductivity of dispersions \hfil (\today)]
{Electrical conductivity of dispersions: from dry foams to dilute
suspensions}

\author{K Feitosa\dag, S Marze\ddag, A Saint-Jalmes\ddag, D J Durian\dag}

\address{\dag\ Department of Physics and Astronomy, University of
Pennsylvania, Philadelphia, PA 19104-6396, USA}

\address{\ddag\ Laboratoire de Physique des Solides, Universit\'e
Paris-Sud, 91405 Orsay, France}

\ead{djdurian@physics.upenn.edu}

\begin{abstract}
We present new data for the electrical conductivity of foams in
which the liquid fraction ranges from two to eighty percent.  We
compare with a comprehensive collection of prior data, and we
model all results with simple empirical formul\ae.  We achieve a
unified description that applies equally to dry foams and
emulsions, where the droplets are highly compressed, as well as to
dilute suspensions of spherical particles, where the particle
separation is large. In the former limit, Lemlich's result is
recovered; in the latter limit, Maxwell's result is recovered.
\end{abstract}

\submitto{\JPCM}

%\pacs{}

 \maketitle

%===============================================================
\section{Introduction}

The electrical conductivity of dispersions is an age-old problem.
One line of research concerns the ``very-wet'' limit, where gas or
liquid bubbles, or spherical solid particles, are widely separated
in a large volume of liquid \cite{MaxwellJC,ClausseEET83}. Another
line of research concerns the ``very-dry'' or ``foam'' limit, where
bubbles are tightly compressed in a small volume of liquid
\cite{Lemlich78, WeaireBook99, WildeCondRev}.  In both, the goal is
to understand the relative conductivity of the dispersion, $\sigma =
\sigma_{sample} / \sigma_{liquid}$, in terms of the volume fraction
of the continuous liquid phase, $\varepsilon$. Experimental
measurement of $\sigma$ could then be used to deduce the value of
$\varepsilon$ for an unknown sample.  The well-accepted behavior is
as follows. In the very-wet limit of $\varepsilon\rightarrow 1$,
Maxwell's result holds: $\sigma = 2\varepsilon/(3-\varepsilon) = 1 -
(3/2)(1-\varepsilon) + (3/4)(1-\varepsilon)^2+
O(1-\varepsilon)^3$~\cite{MaxwellJC}. In the very-dry limit of
$\varepsilon\rightarrow 0$, Lemlich's result holds:
$\sigma=(1/3)\varepsilon$~\cite{Lemlich78}. The former follows from
the form of the electric field in and around an isolated insulating
sphere; the latter follows from the random orientation of Plateau
borders, which are nearly-straight channels of scalloped-triangular
cross-section at which three soap films meet. Considerable effort is
spent on deducing the next-order terms in both wet
\cite{PetersonHermans69, Jeffrey73, TorquatoJAP90, Bonnecaze91,
SalBook} and dry \cite{PhelanJPCM96} limits. Considerable effort is
also spent on developing experimental apparatus for both wet
\cite{Ceccio96, DixonPRB97, Zenit03} and dry \cite{PhelanJPCM96,
Wang99, Cilliers01} extremes. Unfortunately, there is little
understanding of the intermediate regime where both phases occupy
significant volume. Also there are no data sets that span the entire
range of liquid fraction. Furthermore, there appears to be little
contact between researchers focussing separately on the very-wet and
very-dry regimes.  The two lines of research are essentially
disjoint in terms of both theory and experiment.

In this note we explore electrical conductivity in the
intermediate regime.  Our approach is twofold.  First, we scour
the literature for data sets obtained in both wet and dry limits.
Second, we measure the relative conductivity for sequences of
foams with known liquid fraction.  We find that data in the wet
and dry regimes match smoothly, and can be described by simple
empirical formul\ae. This will facilitate experimental studies,
and could guide future theoretical understanding.

\section{Prior observations}

In the very-dry ``foam'' limit, we are aware of three widely-cited
data sets. The first was obtained by Clark for gas bubbles in five
different aqueous solutions \cite{Clark48}.  The second was obtained
by Datye and Lemlich for gas bubbles of different size distributions
in three different aqueous solutions \cite{Datye83}. The third was
obtained by Peters for gas bubbles in aqueous solution
\cite{PetersThesis, PhelanJPCM96}.  A fourth data set was obtained
by Curtayne for two different size bubbles \cite{CurtayneThesis}; a
polynomial fit to this data is shown in Fig.~9.2 of
Ref.~\cite{WeaireBook99}. We extract conductivity data from Fig.~1
of Ref.~\cite{Clark48} and from Fig.~5 of Ref.~\cite{PhelanJPCM96};
A.~Datye kindly provided tables of his data; S.~Hutzler kindly
provided a table of Curtayne's data. In the very-wet limit, we are
aware of three widely-cited data sets. The first was obtained by
Oker-Blom for spherical sand grains set in gelatin, with results
tabulated by Fricke \cite{Fricke24}. The second was obtained by
Meredith and Tobias for oil-in-water emulsions \cite{Meredith61}.
The third was obtained by Turner for solid particles in aqueous
solution \cite{Turner76}. We extract conductivity data from
Table~III of Ref.~\cite{Fricke24}, from Fig.~4 of
Ref.~\cite{Meredith61}, and from Fig.~3 of Ref.~\cite{Turner76}.

The range of data in the dry regime is primarily
$\varepsilon<0.3$, plus two lone points by Clark at
$\varepsilon=\{0.4, 0.54\}$.  The range of data in the wet regime
is primarily $\varepsilon>0.5$, plus one lone point at
$\varepsilon=0.4$ by both Oker-Blom and Turner.  Thus the wet and
dry data sets are nearly non-overlapping, and there is a dearth of
data across the range~$0.3<\varepsilon<0.5$.

\section{New measurements}

To bridge the gap between the data for very-dry and very-wet
regimes, we perform two independent measurements of the relative
conductivity of a sequence of foams.  At Penn, the base aqueous
solution is AOS ($\alpha$-olefin sulfonate, Bio-Terge AS-40 CG-P,
Stepan Company) plus NaCl with concentrations of 8\% and 0.01\% by
weight respectively; the gas is nitrogen.  At Orsay, the base
solution is SDS (sodium dodecylsulfate) plus dodecanol; the gas is
either pure ${\rm C}_2{\rm F}_6$ or else nitrogen plus trace amounts
of ${\rm C}_6{\rm F}_{14}$. In most cases, the foams are produced by
turbulent mixing with an apparatus similar to that of
Ref.~\cite{ArnaudEPJB99}, giving polydisperse bubbles with an
average diameter of 0.1~mm.  For the driest foams at Orsay, bubbles
are created by forcing gas through porous frits; by changing the
porosity, the average bubble diameter can be varied from 1 to 4~mm.

At Penn, foam conductivity is measured as follows.  The foam
delivery hose is connected to an acrylic tube, 30~cm long and
1.27~cm inner diameter, that has brass hose fittings screwed on
both ends. The hose fittings serve as electrodes, which are
connected to an impedance meter (1715 LRC Digibridge, QuadTech).
This meter is configured to measure the resistance of a parallel
resistor-capacitor equivalent circuit, and to operate at a
frequency of 1~kHz and voltage level of 1.00~V.  At this
frequency, the capacitive contribution is negligible. The
resistivity of freshly-produced foam is measured while it flows
downward through the vertically-oriented tube.  The results are
normalized by the resistivity of the base aqueous surfactant
solution, when it entirely fills the tube.  The liquid fraction of
the foam is measured by weighing a known volume of foam, collected
from the output of the acrylic tube concurrently with the
conductivity measurement. The flow speed of the foam is
sufficiently great that no drainage or creaming is observed.

At Orsay, foam conductivity measurements are made in a Plexiglas
column (height 50~cm, and cross section $4\times4~{\rm cm}^2$) in
which 26 pairs of electrodes are embedded, facing each other along
the height \cite{StJalmesJP01}. With this set of electrodes, we
measure the foam conductance with an impedance meter (8284A,
Hewlett-Packard). Frequency and voltage are the same as in the
Penn experiment: 1~kHz and 1~V. For wet foams,
$0.07<\varepsilon<0.50$, the cell is filled with a foam made out
of the turbulent mixer apparatus. The foam conductivity is
measured during the filling and immediately thereafter; absence of
drainage is confirmed by the absence of vertical gradients in
conductivity. For each run, after the cell is filled, a sample of
foam is collected in a calibrated vessel and weighed to determine
liquid fraction.  For dry foams, $0.02<\varepsilon<0.10$, a porous
glass frit is mounted at one end of the conductivity cell, and the
foam is made directly inside it by bubbling gas through the frit,
which is immersed into the surfactant solution. The liquid
fraction is varied by wetting the foam from above with the same
surfactant solution, at a controlled injection rate $Q$. This
method provides uniform foams with no vertical liquid fraction
gradients \cite{StJalmes04}.  The liquid fraction is determined by
measuring the drainage front velocity, $V$, and using the
conservation equation $\varepsilon = Q/(VA)$ where $A$ is the foam
cross section \cite{WeaireBook99}.

Altogether, we have measured relative conductivity over a liquid
fraction range of $0.15<\varepsilon<0.80$ at Penn and
$0.02<\varepsilon<0.50$ at Orsay.

\section{Conductivity vs liquid fraction}

\begin{figure} %-----------------------------------------------
\begin{center}
\includegraphics[width=5.0in]{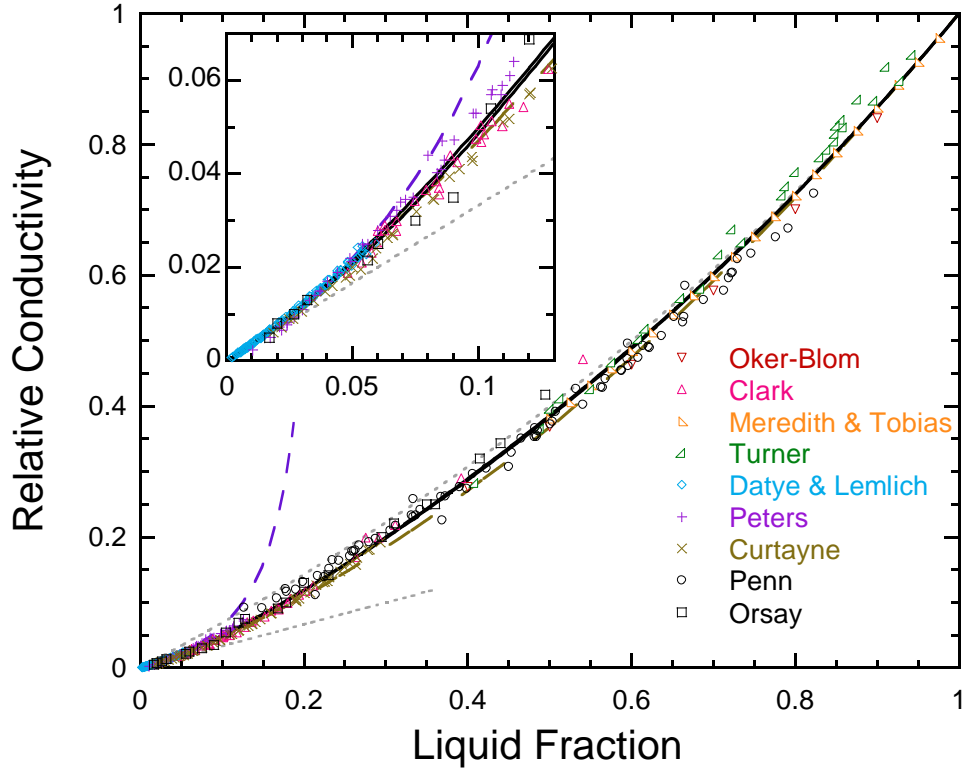}
\caption{Relative conductivity
$\sigma=\sigma_{sample}/\sigma_{liquid}$ vs liquid fraction
$\varepsilon$ for widely-cited data sets, plus our two new data
sets, as labeled. The dotted curves represent the Maxwell and
Lemlich limits, $2\varepsilon/(3-\varepsilon)$ and $\varepsilon/3$,
respectively; the long-dashed curve represent Curtayne's formula,
Eq.~(\protect\ref{Cur}); the short-dashed curve represents the
non-analytic parametric formulation of
Ref.~\protect\cite{PhelanJPCM96}. The solid curves represent
Eqs.~(\protect\ref{SvsE},\protect\ref{EvsS}), which we construct to
have the correct wet and dry limiting behaviors and to fit the data
smoothly in between.} \label{Cond}
\end{center}
\end{figure}

The relative conductivity is plotted vs liquid volume fraction in
Fig.~\ref{Cond} for all data sets, new and old.  As expected, the
Maxwell and Lemlich formul\ae\ appear to hold in their respective
limits. For intermediate liquid fractions, prior very-wet and
very-dry data sets are nearly disjoint but appear to extrapolate
smoothly toward one another. Our new data fill in the gap and bear
this out. This encourages us to seek simple empirical formul\ae\
that hold for {\it all} liquid fraction regimes. We are aware of
three previous suggestions:
\begin{eqnarray}
    \varepsilon &=& 3\sigma - \frac{5}{2}\sigma^{4/3}
        + \frac{1}{2}\sigma^2, \label{Lem} \\
    \sigma &=& \frac{1}{3}\left( \varepsilon + \varepsilon^{3/2}
        + \varepsilon ^2\right ), \label{Cur} \\
    \sigma &=& \frac{1}{3}\varepsilon + \frac{5}{6}\varepsilon^2 -
    \frac{1}{6}\varepsilon^3. \label{WH}
\end{eqnarray}
The first is due to Lemlich \cite{Lemlich85}, the second and third
are due to Curtayne \cite{CurtayneThesis,WeaireBook99}. These three
formul\ae\ all obey the Maxwell and Lemlich limits but underestimate
the conductivity data at intermediate liquid fractions. Curtayne's
Eq.~(\ref{Cur}), plotted as a long-dashed curve in Fig.~\ref{Cond},
comes closer to the data than the other two formul\ae.

 %-------------------------------------------------

Here we suggest modeling the data by rational functions formed by
the ratio of second-order polynomials. From the point of view of a
theorist wishing to predict conductivity in terms of a given liquid
fraction, the appropriate form would be $\sigma =
2\varepsilon(1+A\varepsilon) / [6 + (-7+3A)\varepsilon +
(3-A)\varepsilon^2]$.  From the point of view of an experimentalist
wishing to deduce liquid fraction in terms of the measured
conductivity, the appropriate form would be $\varepsilon =
3\sigma(3+B\sigma) / [3 + (9+2B)\sigma + (-3+B)\sigma^2]$.  All the
numerical coefficients, except for one, are fixed by requiring that
the Maxwell and Lemlich limits be satisfied.  Adjusting the free
parameter to fit the entire collection of data, we find $A=12\pm1$
and $B=33\pm2$. The resulting empirical formul\ae,
\begin{eqnarray}
    \sigma &=& 2\varepsilon(1+12\varepsilon) /
            (6 + 29\varepsilon - 9\varepsilon^2 ), \label{SvsE} \\
    \varepsilon &=& 3\sigma(1+11\sigma)/( 1+ 25\sigma + 10 \sigma^2), \label{EvsS}
\end{eqnarray}
give an excellent description of all data, as shown by the
nearly-identical solid curves in Fig.~\ref{Cond}.  These
formul\ae\ can be used with some confidence owing both to their
agreement with known limits and to the smooth way they
interpolates between data sets in the wet and dry regimes.

Before closing, we compare higher-order behavior with existing
literature. On the wet side, the limiting expansions of
Eqs.~(\ref{SvsE}-\ref{EvsS}) are respectively $\sigma = 1 -
(3/2)(1-\varepsilon) + (0.65\mp0.01)(1-\varepsilon)^2 +
O(1-\varepsilon)^3$. These compare well with
Refs.~\cite{PetersonHermans69, Jeffrey73}, which give the
second-order term as $0.656(1-\varepsilon)^2$ and
$0.588(1-\varepsilon)^2$ respectively.  On the dry side, the
limiting expansions of Eqs.~(\ref{SvsE}-\ref{EvsS}) are respectively
$\sigma = \varepsilon/3 + (2.0\pm0.4)\varepsilon^2 +
O(\varepsilon)^3$. These cannot be directly compared with
Ref.~\cite{PhelanJPCM96}, which proposes a non-analytic parametric
formulation of liquid fraction and relative conductivity as
$\varepsilon = 0.171 a^2 (1+1.5a)$ and $\sigma = 0.171
a^2/(3-3.81a)$; eliminating the parameter $a$ and expanding gives
$\sigma = \varepsilon/3 - 0.185\varepsilon^{3/2} + 4.15\varepsilon^2
- 19.6\varepsilon^{5/2}+\ldots$.  On this basis, we attempt to
describe the data by $\sigma = [\varepsilon +
(1+C)\varepsilon^{3/2}] / [3 - (1-2C)\varepsilon^{1/2} -
C\varepsilon]$, which also obeys the Maxwell and Lemlich limits. The
best fit, $C=2.8\pm0.8$, has a $\chi^2$ value that is about ten
percent worse than that for Eq.~(\ref{SvsE}); it gives the leading
correction to Lemlich as $+0.76\varepsilon^{3/2}-0.85\varepsilon^2$.
Due to scatter in the data (see inset of Fig.~\ref{Cond}), we cannot
rule out either of these contrasting non-analytic expansions.

\section*{Acknowledgments} We thank J.F.~Brady, S.~Hutzler,
D.L.~Koch, S.~Torquato, and D.~Weaire for helpful discussions, and
we thank A.~Datye and S.~Hutzler for kindly providing tables of
data. This material is based upon work supported by NASA
Microgravity Fluid Physics under grant NAG3-2481.

%=====================================================================

\end{document}